    \documentclass[12pt]{article} % \usepackage{dina4p} 
    \usepackage{psfig}  
    \usepackage{epsfig}  
    \usepackage{graphics}
    \usepackage{subfigure}
    \usepackage{graphicx}
     \newlength{\dinwidth}                       
     \newlength{\dinmargin}                      
\def\Journal#1#2#3#4{{#1} {\bf #2}, #3 (#4)}

\def\NPB{{\em Nucl. Phys.} B}
\def\PLB{{\em Phys. Lett.}  B}

\def\PRD{{\em Phys. Rev.} D}
\def\ZPC{{\em Z. Phys.} C}
\def\EPC{{\em Eur. Phys. J.} C}  % \em added !!
\def\lsim{\mathrel{\rlap{\lower4pt\hbox{\hskip1pt$\sim$}}
    \raise1pt\hbox{$<$}}}                % less than or approx. symbol
\def\gsim{\mathrel{\rlap{\lower4pt\hbox{\hskip1pt$\sim$}}
    \raise1pt\hbox{$>$}}}                % greater than or approx. symbol
\newcommand{\eg}{{\it e.g.\ }}
\newcommand{\ie}{{\it i.e.\ }}

\newcommand{\ket}{\rangle}
\begin{document}
\thispagestyle{empty}

%\begin{document}
\mbox{ }
\vspace{20mm}

\includegraphics{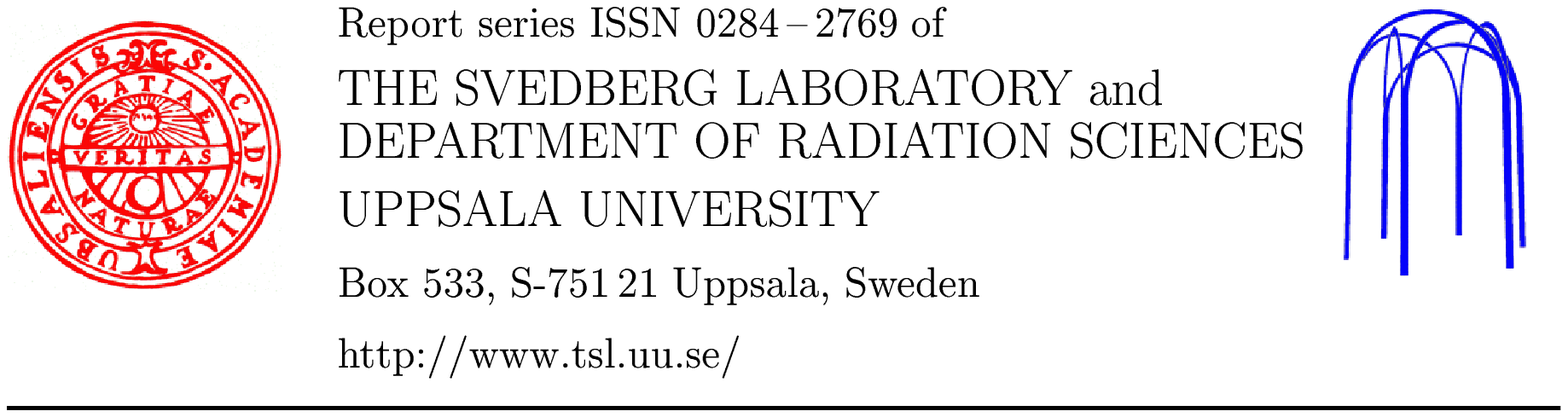}

\begin{flushright}
\begin{minipage}[t]{37mm}
{\bf TSL/ISV-95-0214 \\
August 1999} 
\end{minipage}
\end{flushright}

\begin{center}
\vspace{5mm}

{\bf \Large Non-perturbative Parton Distributions in the Proton\footnote{In proceedings 
`Monte Carlo generators for HERA physics', DESY-PROC-1999-02, 
www.desy.de/\~{}heramc}}

\vspace{20mm}

{\Large A.~Edin$^a$, G.~Ingelman$^{ab}$, K. Torokoff$^b$} \\
\vspace{3mm}
$^a$ Deutsches Elektronen-Synchrotron DESY, Hamburg, Germany\\
$^b$ High Energy Physics, Uppsala University, Sweden, www3.tsl.uu.se/thep\\

\end{center}

\vspace{5mm}

{\bf Abstract:} 
A model for the parton distributions in hadrons is derived from simple physical
arguments leading to an analytical expression for the valence distributions. 
The sea parton distributions arise mainly from pions in hadronic fluctuations.
Evolving from a low $Q^2_0$ scale with DGLAP gives the proton structure 
function $F_2(x,Q^2)$ in good agreement with DIS data. An extension of the 
model to describe intrinsic charm quarks in the proton is discussed. 

\vfill
\clearpage
\setcounter{page}{1}
\noindent
TSL/ISV-99-0214    \hfill ISSN 0284-2769\\
August 1999        %  \hfill {\bf DRAFT} \today\\
\vspace*{5mm}
%\vspace*{10mm}
\begin{center}  \begin{Large} \begin{bf}
Non-perturbative Parton Distributions in the Proton\footnote{In proceedings 
`Monte Carlo generators for HERA physics', DESY-PROC-1999-02, 
www.desy.de/\~{}heramc}
\\
  \end{bf}  \end{Large}
%  \vspace*{5mm}
  \begin{large}
A.~Edin$^a$, G.~Ingelman$^{ab}$, K. Torokoff$^b$\\
  \end{large}
\end{center}
$^a$ Deutsches Elektronen-Synchrotron DESY, Hamburg, Germany\\
$^b$ High Energy Physics, Uppsala University, Sweden, www3.tsl.uu.se/thep\\
\begin{quotation}
\noindent
{\bf Abstract:}
A model for the parton distributions in hadrons is derived from simple physical
arguments leading to an analytical expression for the valence distributions. 
The sea parton distributions arise mainly from pions in hadronic fluctuations.
Evolving from a low $Q^2_0$ scale with DGLAP gives the proton structure 
function $F_2(x,Q^2)$ in good agreement with DIS data. An extension of the 
model to describe intrinsic charm quarks in the proton is discussed. 
\end{quotation}

\section{Introduction}

When calculating cross sections for hard processes with incoming hadrons, the
structure of the hadrons is described by parton distributions. The dependence
of the parton distributions on the hard scale $Q$ of the interaction is
described by perturbative QCD evolution. However, their dependence on the
momentum fraction $x$ at the lower limit for applying perturbative QCD,
$Q_0\approx 0.5-2$ GeV, cannot be calculated perturbatively and must therefore
be obtained by other means. 

The easiest solution is to fit a parameterization~\cite{CTEQ5,GRV98,MRS}, \eg
of the form
$f_i(x,Q_0)=N_i x^{a_i} (1-x)^{b_i} (1+c_i \sqrt{x} + d_i x)$, 
to experimental data and then use these functions when calculating cross
sections. The problem with this approach is that the  parameters in these
functions have no direct physical meaning, making it difficult to interpret the
results. 

To gain understanding of non-perturbative QCD and the structure of hadrons we
have developed a physical model \cite{EI_partons98,EI_dis99} for the parton
distributions at $Q_0$. 
The model is here briefly described together with our latest developments. 

\section{The model}

\begin{figure}[tb]
\center{
\subfigure[ ]{\includegraphics[width=50mm]{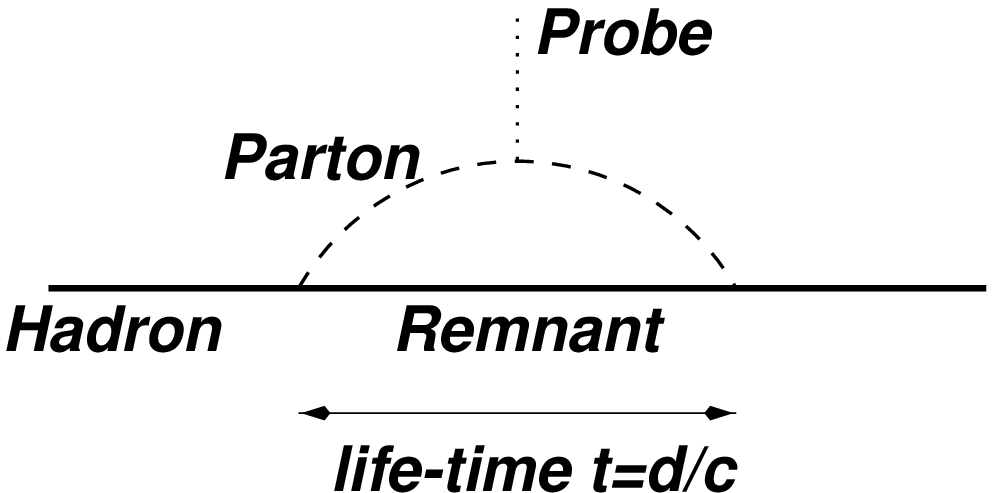}}
\subfigure[ ]{\includegraphics[width=55mm]{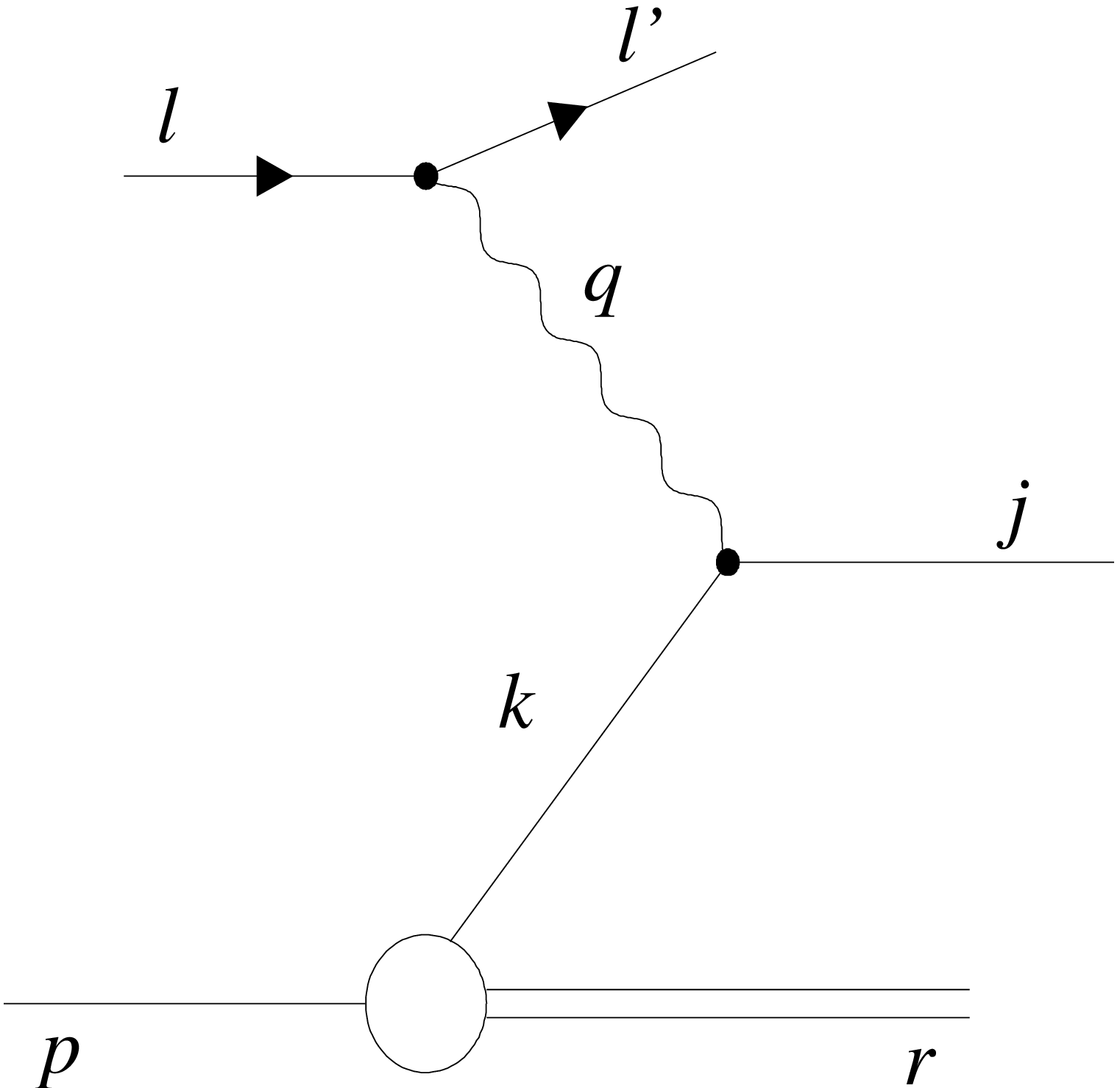}}
\subfigure[ ]{\includegraphics[width=50mm]{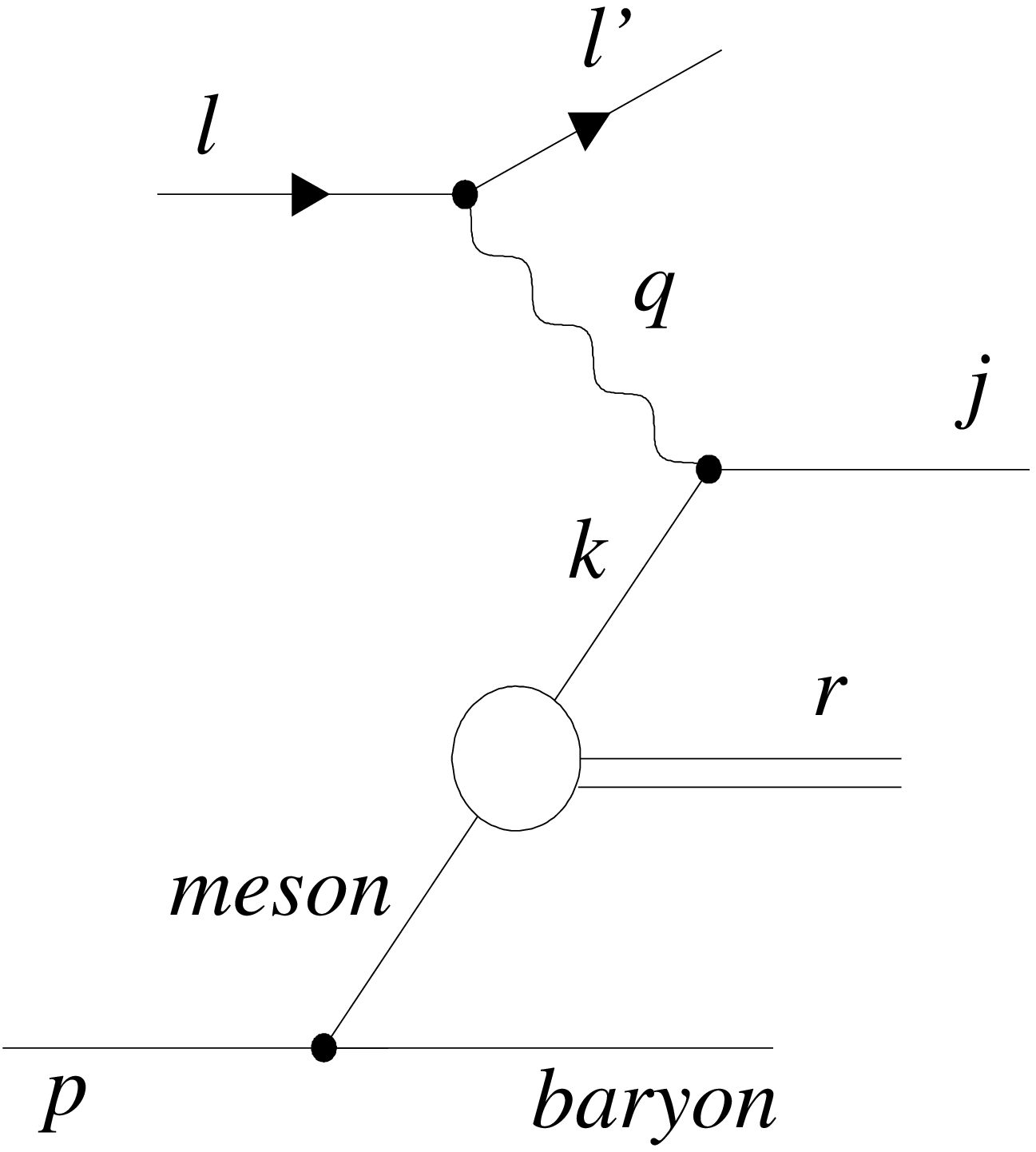}}
}
\caption{
(a) Time scales in probing a parton in a hadron. 
Kinematics and four-vectors for scattering on 
(a) a valence parton and (b) a sea parton in a hadronic fluctuation. 
}
\label{fig:kin}
\end{figure}

The basic assumption in the model is that the probe resolving the hadron has
much higher resolution than the size of the hadron. The probe will then see
{\it free quarks and gluons} in quantum fluctuations of the hadron. 
As illustrated in Fig.~\ref{fig:kin}a, the measuring time is short compared 
to the life-time of the fluctuation which is given by the confinement scale. 
This makes it possible to describe the formation of the fluctuation 
independently of the measuring process. 

Our approach only intends to provide the probability distribution of the 
four-momentum $k$ of a single probed parton. 
All other information in the hadron wave function is neglected,
treating the other partons collectively as a remnant with four-momentum $r$,
see Fig.~\ref{fig:kin}b. The parton distribution is described using the
light-cone momentum fraction  $x=\frac{k_+}{p_+}$ which the parton has in the
initial hadron. Since $x$ is invariant under boosts in the $z$ direction, the
same will be true for the calculated parton distributions. 

It is most convenient to describe the fluctuations in the hadron rest frame 
where there is no preferred direction and hence spherical symmetry. The momentum
distribution of the partons should therefore be spherically symmetric. 
The shape of the momentum distribution should be close to a Gaussian as a 
result of the many interactions binding the parton in the hadron. The typical 
width of this distribution should be a few hundred MeV from the Heisenberg
uncertainty relation applied to the hadron size. 
The Gaussian momentum distribution also has phenomenological support. 
The Fermi motion in the proton provides the `primordial transverse momentum',
which has been extracted from DIS data and found to be well described by a 
Gaussian distribution of a few hundred MeV width \cite{kt}. 

\subsection{Valence partons}
The probability distribution in momentum space for finding one parton $i$ 
with mass $m_i$ is therefore give by the Gaussian
\begin{equation}
f_i(k)=N(\sigma_i,m_i)
          e^{-\frac{(k_0-m_i)^{2}+k_1^2+k_2^2+k_3^2}{2\sigma_i^2}},
\label{eq:fk}
\end{equation}
where $\sigma=\frac{1}{d_h}\approx m_{\pi}$ is the inverse of the confinement 
length scale, \ie the hadron diameter. 
The Gaussian in the energy component comes from the limited time the
parton can be `free'.

The parton distributions must fulfill a number of constraints. The
normalization for valence quarks is given by the sum rules and for the gluons
by the momentum sum rule, \ie 
\begin{equation}
\int_0^1 f_i(x)dx = n_i \;\;\; {\rm and} \;\;\;
\sum_i\int_0^1 xf_{i}(x)dx = 1.
\end{equation}
There are also the kinematic constraints 
$m_{i}^{2}\le j^{2}<W^{2}$ and $r^{2}>\sum _{i}m_{i}^{2}$
given by the final parton $j$ being 
on-shell or time-like and the remnant having to include the remaining partons. 
This also leads to $0<x<1$. 

The parton model requires that $W$ is well above the resonance region and that 
the resolution of the probe is much larger than the size of the hadron, \ie 
$W \gg m_{p} \;\;\; {\rm and} \;\;\; Q_0 \gg \sigma _{i}$. 
The scale of the probe must also be large enough, $Q_0\gg \Lambda _{QCD}$,  for
perturbative QCD to describe the evolution of the parton distributions from the
starting scale $Q_0$.

In~\cite{EI_partons98} we integrated eq.~\ref{eq:fk} numerically 
to find the parton distributions since the kinematic 
constraints are quite complicated in general. 
The problem is much simpler if the transverse momenta and the masses of the
partons are neglected. It is then possible to derive \cite{E_partons99}
an analytical expression for the parton distributions,  
\begin{equation}
f_i(x)=N'(\tilde{\sigma}_i)\exp \left( -\frac{x^2}{4\tilde{\sigma}_i^2}\right) 
{\rm erf}\left( \frac{1-x}{2\tilde{\sigma}_i}\right)
\;\;\; {\rm where} \;\;\; 
\tilde\sigma_i = \frac{\sigma_i}{m_h} =  
\frac{1}{d_h m_h} \approx \frac{m_{\pi}}{m_h}.
\label{eq:fx}
\end{equation}
The valence parton distributions for hadrons are here determined simply by 
the mass and size of the hadron! 

\begin{figure}[hbt]
\vspace*{10mm}
\center{
\includegraphics[width=75mm]{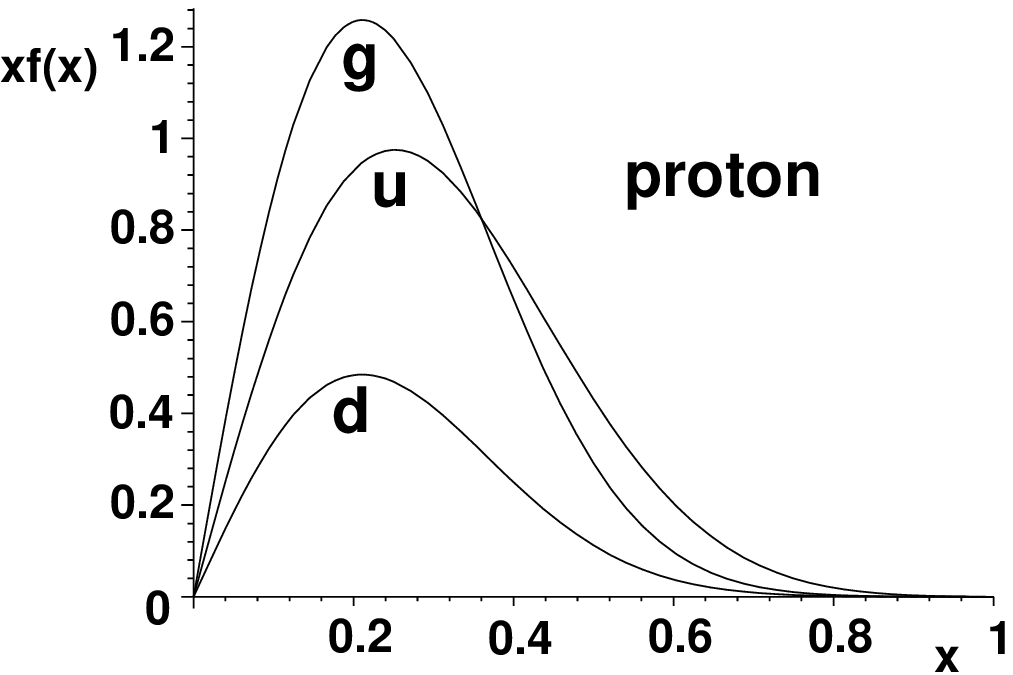}
\hspace{10mm}
\includegraphics[width=75mm]{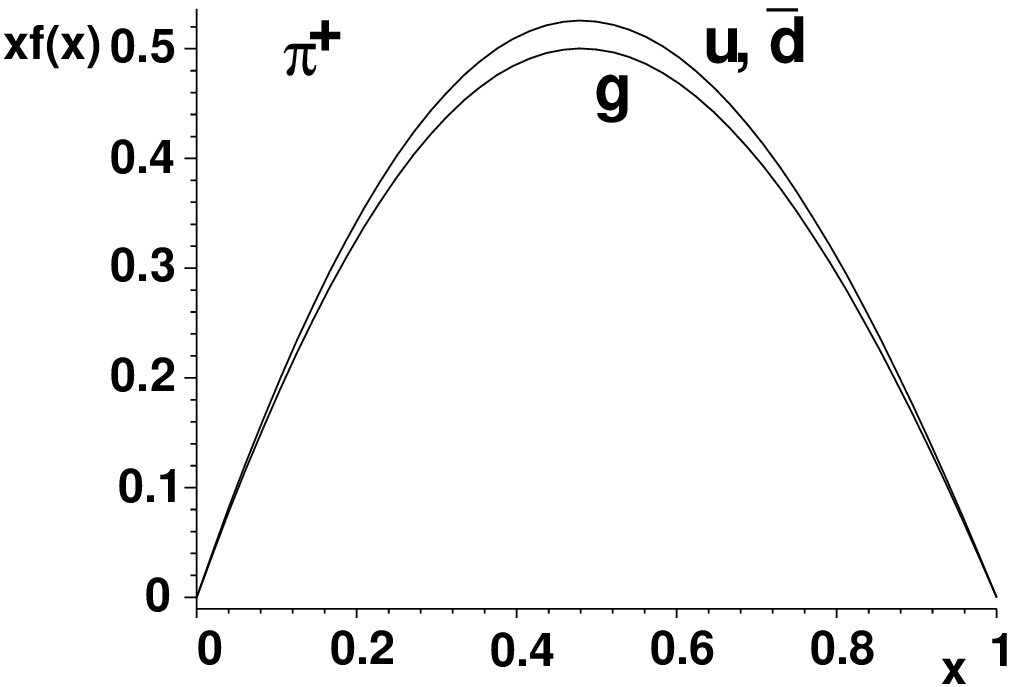}
}
\caption{Valence distributions at $Q_0\approx 0.85$ GeV in the proton and 
$\pi^+$.} 
\label{fig:proton_pion}
\end{figure}

The resulting valence distributions for the proton ($\tilde\sigma\approx 0.15$)
and the pion ($\tilde\sigma\approx 1$)
are very reasonable as shown in Fig.~\ref{fig:proton_pion}. 
Note that the pion distributions are very similar to $xf(x)=2x(1-x)$ and that 
one third of the pion momentum is carried by gluons.

\subsection{Sea partons}
The sea partons are described by hadronic fluctuations as illustrated in
Fig.~\ref{fig:kin}c. For example, the proton fluctuates into 
$|p\pi^0\ket+|n\pi^+\ket+...$ and the scattering is on a valence parton in
one of the two hadrons, most importantly the pion. The sea quark momentum
distribution is obtained as the convolution of the valence parton in the pion, 
which was already obtained from the model, with the momentum distribution 
of the pion in the hadronic fluctuation. 
The latter is assumed to follow from the same model, \ie using a spherically 
symmetric Gaussian momentum distribution in the proton rest frame. However, 
the width $\sigma_{\pi}\approx 50$ MeV is smaller, related to the longer 
range of pionic strong interactions or the size of the virtual pion cloud 
around a proton.

\section{Proton structure function and DIS data}

\begin{figure*}[hbt]
\center{
\includegraphics[height=65mm]{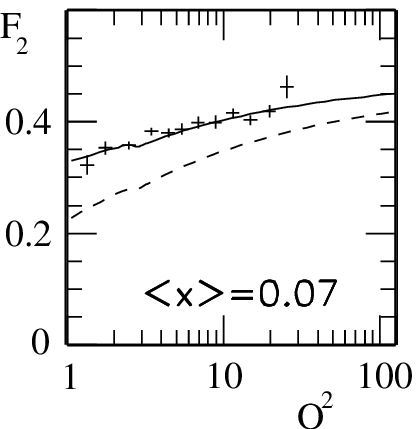}
\hspace{10mm}
\includegraphics[height=65mm]{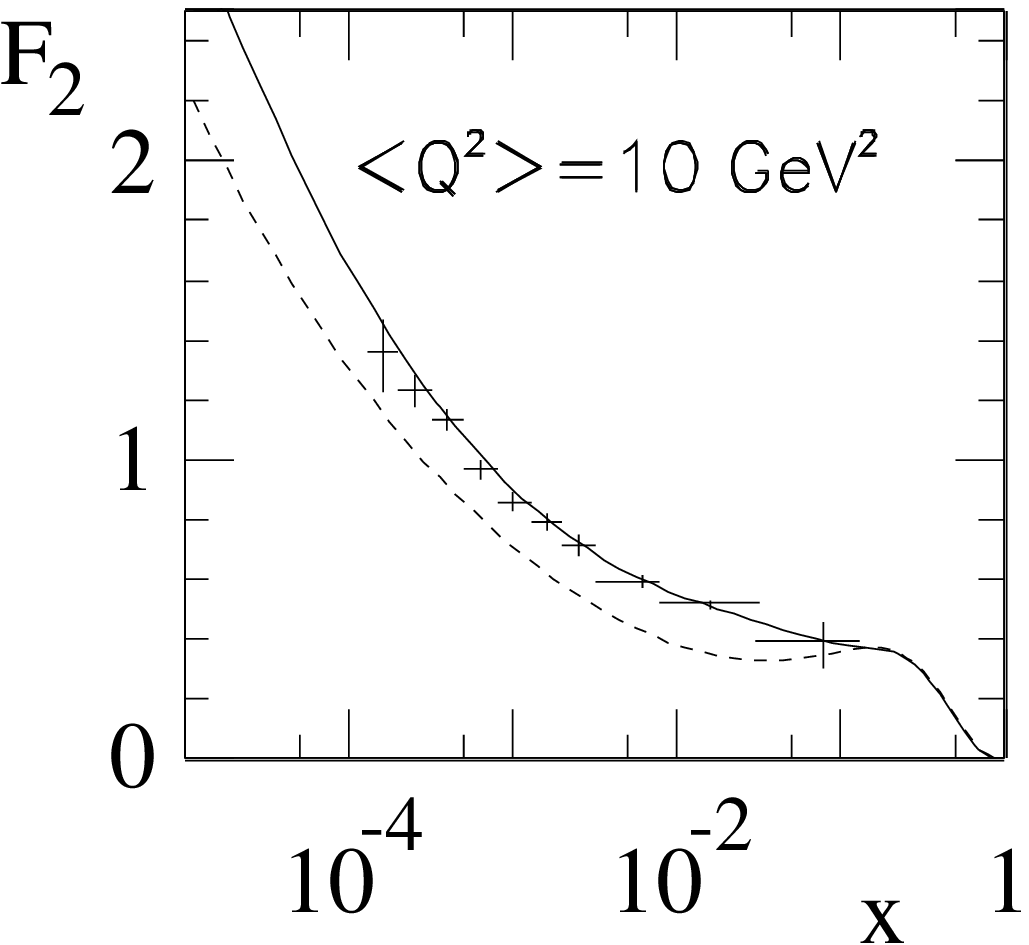}
}
\caption{Data from NMC \cite{NMC} (left) and ZEUS \cite{ZEUS} (right) on the 
proton structure function $F_2(x,Q^2)$ compared to the model with valence 
partons only (dashed) and including sea partons (full). 
Extracted from \cite{EI_partons98} where larger $x,Q^2$ regions are examined.}
\label{fig:nmc}
\end{figure*}

Using these valence and sea parton $x$ distributions at $Q_0$, next-to-leading 
order DGLAP evolution in the CTEQ computer program \cite{CTEQ5} was
applied to obtain the parton distributions at larger $Q$. The proton structure
function $F_2(x,Q^2)$ was then calculated and compared with deep inelastic
scattering data as illustrated in Fig.~\ref{fig:nmc}. 
The model curves were obtained \cite{EI_partons98}
using the full numerical integration including effects from masses, transverse 
momenta and exact kinematic constraints rather than with the simplified 
analytical expression in eq.~(\ref{eq:fx}). Furthermore, the model parameters
are here fitted to the data giving $Q_0=0.85$ GeV, 
$\sigma_u=180$ MeV, $\sigma_d=150$ MeV, $\sigma_g=135$ MeV, 
$\sigma_{\pi}=52$ MeV and 7.7\% of the proton momentum 
carried by sea partons (\ie pion fluctuation normalization).
These values are in very good agreement with the definite expectations of 
the model. In particular, the values $\sigma_i$ in eq.~(\ref{eq:fk}) 
corresponds to the Fermi motion in the proton. (Since $\sigma$ applies in each
dimension, one obtains a two-dimensional distribution with 
$\langle k_\perp^2 \rangle=2\sigma^2$ in agreement with data on the primordial
transverse momentum.) One may note that that the $u$-quark distribution has 
a $\sim 20\%$ larger width, which may correspond to a correspondingly smaller 
available region due to the Pauli exclusion principle. 

The model does remarkably well in view of its simplicity and few parameters (6).
Of course, conventional parton density parameterizations give better fits, 
probably mainly due to their many more parameters ($\approx 20$). 
The main advantage with our approach is that it provides a model that 
not only gives a good representation of data, but also provides new insights
through a physical model for the non-perturbative input distributions. 

The main part of the proton structure is determined by the valence
distributions, but the sea gives an important contribution at small $x$. 
For $Q^2 < 1$ GeV$^2$ the DIS formalism 
and the partonic interpretation of $F_2$ are not straightforwardly applicable. 
Photoproduction and resolved photon processes must here be included in
the theoretical description of the data.
More generally, parton distributions can only be used 
when the hadron is resolved by a sufficiently large scale. 
Hence, for $Q^2<Q^2_0$ our parton distribution model can only be applied 
for processes with some other hard scale, \eg jets in photoproduction, 
and not to obtain the total photon-proton cross section or $F_2$ extracted 
from it. 

\section{Intrinsic charm}
Following the model described above we have made a detailed investigation 
\cite{K_Xjobb} on how it can be extended to describe intrinsic charm quarks 
in the proton. The main results and their comparison with the original 
intrinsic charm model of Brodsky el al. \cite{Brodsky} are briefly discussed
here. Studies on the possibility to observe intrinsic charm in DIS have 
been presented earlier \cite{IC-search}.

The idea that our model for parton distributions may be applied to describe 
intrinsic charm is obvious, but there are problems of both conceptual and 
technical nature. Extending the model for generating sea quarks through 
hadronic fluctuations, illustrated in Fig.~\ref{fig:kin}c, means that one 
should consider fluctuations 
$|p\rangle \to  |\Lambda_c^+\overline{D}^0\rangle + \ldots$. 
In old-fashioned perturbation theory the particles are on-shell and the 
fluctuation is suppressed by a factor proportional to $1/\Delta E^2$.
This provides a description with energy fluctuation in the hadron basis 
(EFH). 

The large mass of the charm quark and charm hadrons means a large fluctuation
$\Delta E$ and one could argue that the fluctuation is correspondingly 
short-lived and that the hadron basis is not suitable since hadron states 
are not be formed. One should then picture the charm quark as coming
directly from the proton, as in Fig.~\ref{fig:kin}b, with a remnant containing 
the charm anti-quark (or vice versa) to conserve the charm quantum number. 
This gives a description with energy fluctuation in the parton basis (EFP). 

An alternative description is to impose energy-momentum conservation, which 
means that in the hadron (parton) basis the charm meson (quark) is off-shell.
The charm baryon (remnant) is external to the hard scattering process
and should therefore be on-shell. (The remnant can be slightly off-shell if 
one allows it to pick up some energy in the hadronisation process.) 
This provides two model alternatives with energy conservation in the hadron 
or parton basis (ECH and ECP). 

\begin{figure}[hbt]
\vspace*{-10mm}
\center{
\includegraphics[width=85mm]{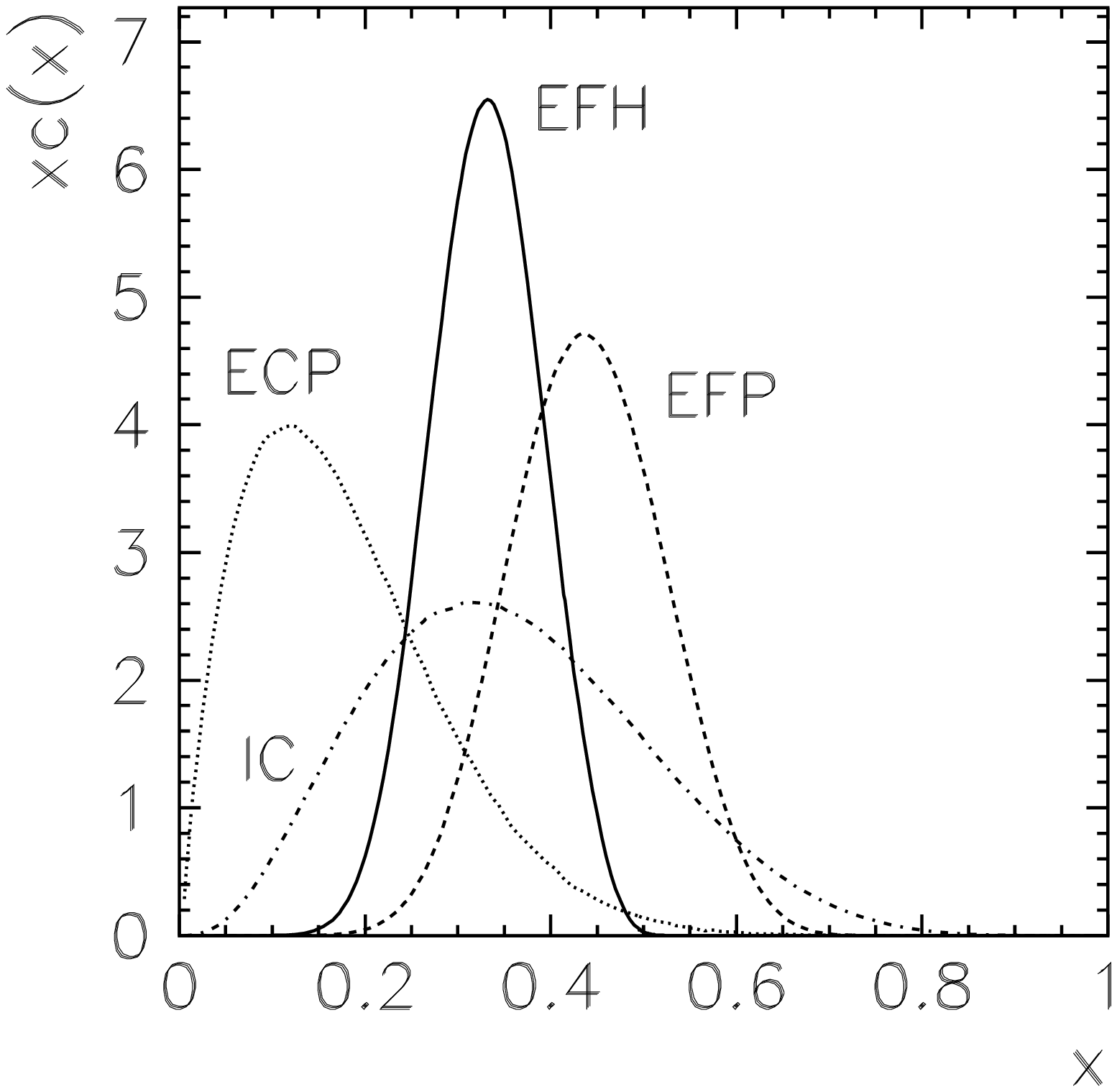}
\hspace*{-5mm}
\includegraphics[width=85mm]{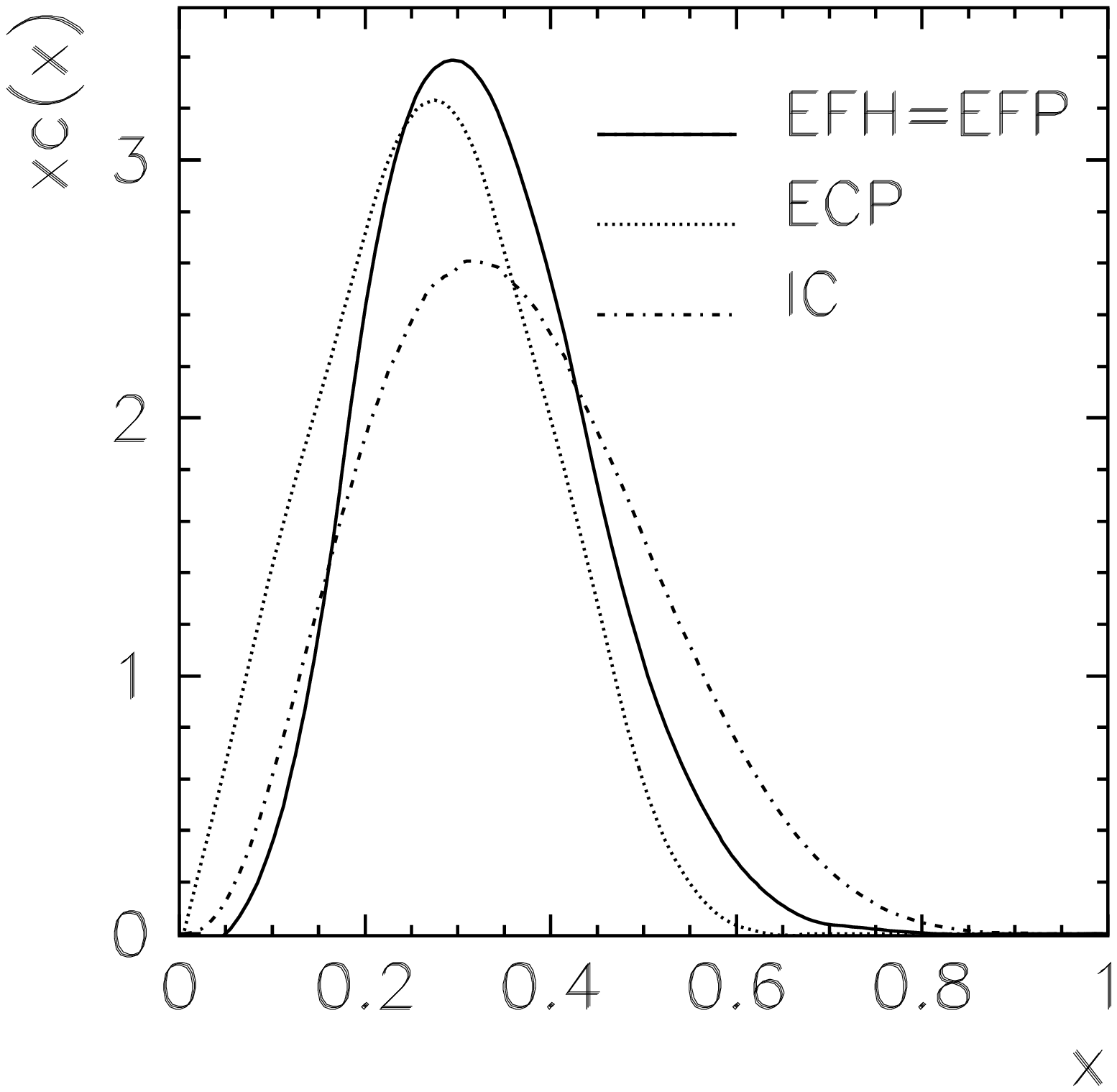}
}
\vspace*{-5mm}
\caption{
Momentum-weighted distributions $xc(x)$ for intrinsic charm in the proton
obtained from variations of our model (EFH and EFP with energy fluctuation 
in hadron and parton basis, respectively, and ECP with energy conservation 
in parton basis) with (left) and without (right) taking into account kinematic 
constraints. Comparison with the original intrinsic charm (IC) model of 
Brodsky et al.\ (without constraints). All curves are normalized to unity area.} 
\label{fig:IC}
\end{figure}

The kinematic constraints are imposed as previously, but they become harder
due to the large charm mass. In fact, the ECH model is then not kinematically
allowed. The results of the other model variations are shown in 
Fig.~\ref{fig:IC}, with and without the kinematic constraints applied. 
In case of fluctuations in a parton basis there is quark-anti-quark symmetry,
$xc(x)=x\bar{c}(x)$, while for hadronic fluctuations this is not the case
since the $\overline{D}$ meson is softer than the $\Lambda_{c}$ baryon due 
to its lower mass.

As shown in Fig.~\ref{fig:IC}a, the different model variations result in 
somewhat different charm distributions. However, if the kinematic constraints
are not applied, then they give essentially the same $x$-distributions 
as demonstrated in Fig.~\ref{fig:IC}b. They are also then quite similar to 
the $x$-distribution in the original intrinsic charm model \cite{Brodsky}
where these kinematic constraints are not accounted for. The sensitivity of
our results on the treatment of the charm quark mass and the constraints 
demonstrate the importance of taking them properly into account to ensure 
a kinematically allowed final state. 

The results of the models have been studied in more detail \cite{K_Xjobb} 
by examining the four-vectors in the process. As expected, the incoming 
quark $k^2$ is almost on-shell in the energy fluctuation schemes, 
while in the energy conserving scheme it is off-shell and negative.
In ECP, the $j^2$ distribution peaks at $m_c^2$ and falls very quickly for 
larger values. In EFP and EFH, however, $j^2$ is larger. This may only 
be an artifact which can be avoided in the model, but it can also signal that 
perturbative gluon emission must be taken into account. If so, also 
boson-gluon fusion should be included for consistency and then perturbative 
charm production would enter such that one would have to define a suitable
factorization to disentangle the non-perturbative intrinsic charm component.

Our results here and in \cite{K_Xjobb} are therefore only a first step towards
a realistic model for intrinsic charm based on our general idea for deriving
parton distributions. 

\section{Conclusions and outlook}
We have shown that our new model for parton momentum distributions 
in hadrons is reasonable and can reproduce the measured proton structure
function. They can therefore be used as an alternative to the conventional 
parameterizations in practical calculations of hard processes and in 
Monte Carlo generators. The advantage of the physical model over the 
parameterizations is that it provides insights on the non-perturbative 
dynamics embodied in the model. 

The experimentally observed \cite{Sea_asymmetry} flav\-our asymmetric sea with 
$\bar{d}>\bar{u}$ may be accounted for by including all pion fluctuations, 
but the numerical details remain to be investigated. The hadronic fluctuations 
may also be used to develop a model for the dynamics of leading baryon 
production ({\it cf.} Fig.~\ref{fig:kin}c). 

The application of the model to other hadrons than the proton will be 
presented elsewhere \cite{E_partons99}, but the results cannot be as directly
and precisely tested as in deep inelastic scattering on the nucleon. 
However, data on the pion structure are available from Drell-Yan processes 
in pion beam experiments.

\newpage
Applying the model to charm in the proton is interesting, but more complicated
due to the large charm mass. Although reasonable results 
are obtained, the $x$-distribution depends on the details in the treatment of
the charm quark mass effects and the related kinematic constraints imposed 
to ensure an allowed final state. The proton remnant system plays an important
role in this respect such that its dynamics may be essential for developing a 
realistic model.

\end{document}